\documentclass[]{spie}  

 \usepackage{booktabs}
\usepackage{amsmath,amsfonts,amssymb}
\usepackage{graphicx}
\usepackage[colorlinks=true, allcolors=blue]{hyperref}
\usepackage{caption}
\title{A UNIFIED DEEP LEARNING FRAMEWORK FOR CONTRAST-PHASE-SPECIFIC VIRTUAL MONOCHROMATIC IMAGING}

\author[a]{Antony Jerald}
\author[a]{Hemant K Aggarwal}
\author[b]{Brian Nett}
\author[c]{Avinash Gopal}
\author[d]{Phaneendra K Yalavarthy}
\author[a]{Bipul Das}
\author[a]{Rajesh Langoju}
\affil[a]{Science and Technology Organization, GE HealthCare, Bangalore, INDIA}
\affil[b]{CT Engineering, GE HealthCare, Waukesha, USA}
\affil[c]{Science and Technology Organization, GE HealthCare, San Ramon, USA}
\affil[d]{Medical Imaging Group, Dept. of Computation and Decision Sciences, Indian Institute of Science, Bangalore, INDIA}

\authorinfo{Author of correspondence: rajesh.langoju@gehealthcare.com }
\begin{document} 
\maketitle

\begin{abstract}
Dual-energy CT (DECT) enables virtual monochromatic imaging (VMI) and improved contrast resolution, but its clinical adoption is limited by hardware complexity and cost. In this work, we propose a unified deep learning framework that synthesizes contrast-phase-specific virtual monochromatic 50~keV images from single-energy CT (SECT) data by leveraging contrast phase information as a prior. The model is trained using DECT-derived 70~keV and 50~keV image pairs across four contrast phases - Angio, Arterial, Portal, and Delayed - using a novel prior conditioning architecture that integrates contrast phase priors into the energy transformation process. We demonstrate that the proposed unified model achieves contrast enhancement and generalizes well across contrast phases. Additionally, we show that the model can generate 50~keV-like images from SECT inputs, preserving contrast phase-specific dynamics.
\end{abstract}

\keywords{Contrast CT Imaging, Spectral CT, Virtual Monochromatic Imaging, Multi-task Learning}

\section{INTRODUCTION}
Contrast-enhanced CT (CECT) is a diagnostic imaging technique that uses intravenous (IV) injection of iodine-based contrast agents and timed imaging phases to visualize specific anatomical structures and functions. After IV contrast administration, images are captured at defined intervals: the Angio phase (15-20~s) highlights arteries for vascular assessments; the Arterial phase (25-35~s) captures arteries and early portal vein enhancement, useful for hypervascular tumours; the Portal venous phase (60-90~s) focuses on liver and abdominal organs; the Delayed phase (3-5~mins) is useful for liver lesions and renal evaluation. Multi-phase imaging enables targeted diagnostic studies including oncology analysis and follow-up in liver, pancreas, and other anatomies. Dual-Energy CT (DECT) is preferred over Single-Energy CT (SECT) for contrast studies due to its ability to acquire data at two energy levels, enabling virtual monoenergetic imaging (40-140~keV), improved iodine visualization, leading to better analysis of vasculature and tissue permeability for oncological studies, and reduced radiation exposure through virtual non-contrast imaging~\cite{johnson2012dual, hamid2021clinical}. However, its adoption is limited by cost, complexity, and patient-specific challenges.

Recent advances in deep learning (DL) offer a promising alternative for synthesizing virtual monochromatic images from single-energy CT (SECT) data by learning non-linear mappings. Prior studies have shown the feasibility of generating virtual monochromatic (VM) images using convolutional neural networks and hybrid architectures \cite{lyu2021estimating,cong2020virtual,zhao2020deep,fink2022jointly}. However, transforming SECT images (e.g., 120 kVp) into low-keV representations (e.g., 40 or 50 keV like) is sensitive to the contrast phase timeline, as different anatomical regions retain contrast differently across phases. This necessitates phase-specific models for accurate energy transformation. In this work, we propose a unified framework for contrast phase-specific VM imaging, where a single model is trained to perform energy transformation across multiple contrast phases—Angio, Arterial, Portal, and Delayed—using contrast phase information as a prior. The model is trained jointly on GE HealthCare, Gemstone Spectral Imaging (GSI) mono-keV data (e.g., 70 keV and 50 keV) from all phases. The known contrast phase timeline is converted into a unique code, processed through a conditioning network and finally integrated into the main network to guide the transformation process. Details on data preparation, and training methodology are provided in the methods section.

\section{METHOD}
\begin{figure}[htbp]
	\centering
	\includegraphics[width=0.99\textwidth]{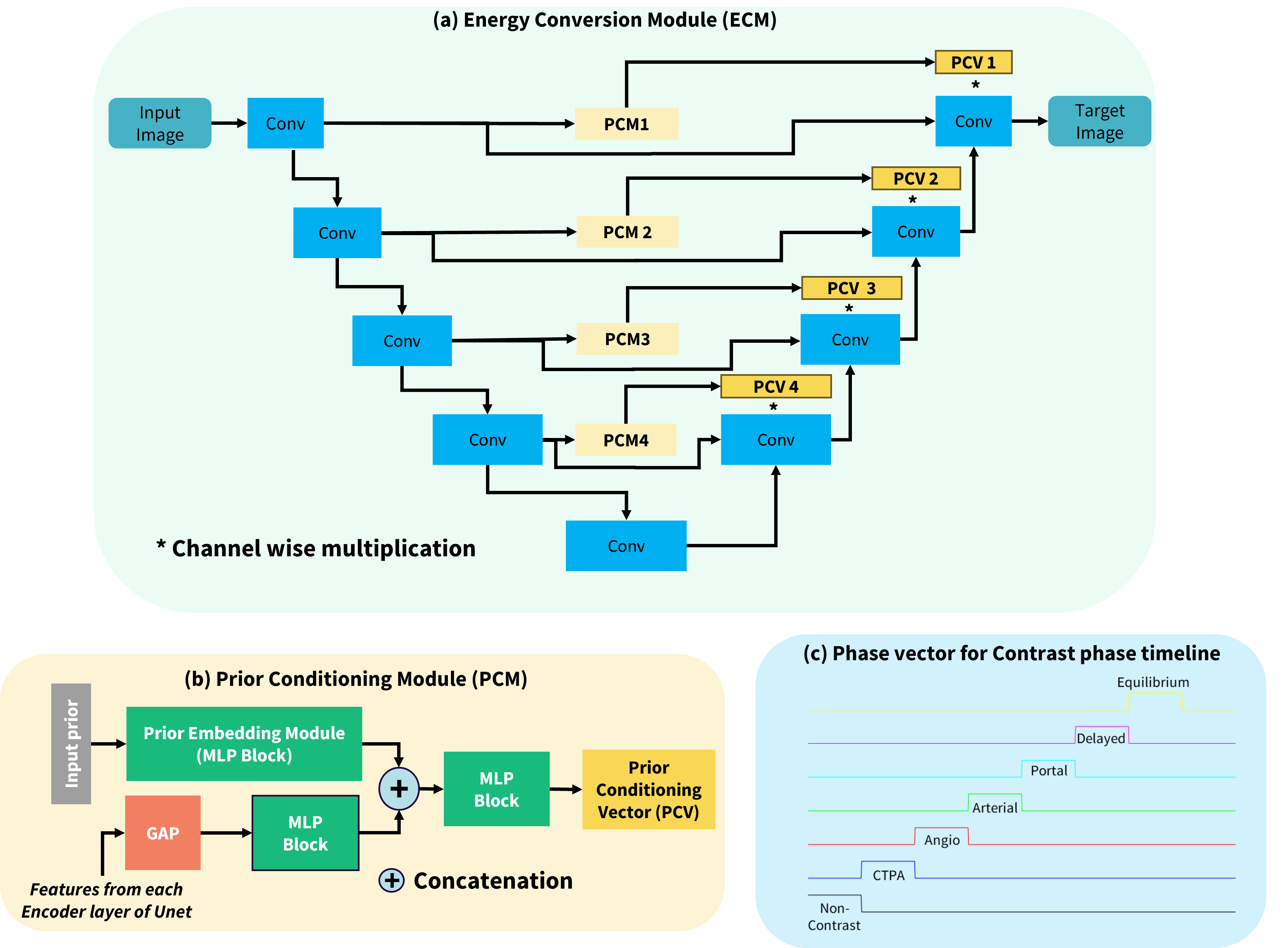}
	\caption{(a) Overall Architecture for energy transformation across phase-contrasts used in the proposed approach. (b) Proposed prior conditioning module used for adaptively modulate decoded features based on the input contrast phase. (c) Contrast phase timelines used as prior information  }
	\label{fig:Arc}
\end{figure}

We propose a supervised learning approach for energy transformation in CT imaging. Since single-energy CT (SECT) images acquired at 120 kVp lack corresponding low-keV target images, so our method first learns the transformation using dual-energy CT (DECT) data and then applies it to SECT inputs. Given that 120 kVp images are comparable to 70 keV in terms of tissue attenuation \cite{matsumoto2011virtual,krishna2018attenuation,cui2012should}, we train the model using 70 keV and 50keV image pairs from DECT system as input and target respectively.  The energy transformation architecture, illustrated in Fig. \ref{fig:Arc}, consists of two modules: (1) the Energy Conversion Module (ECM) shown in Fig. \ref{fig:Arc} (a) which is  a standard U-Net architecture \cite{ronneberger2015u} with four downsampling and upsampling stages, which is the primary network that is responsible for converting 70 keV images into 50 keV images. (2) the Prior Conditioning Module (PCM) shown in Fig. \ref{fig:Arc} (b), that encapsulates phase information of the given scan and converts this information into prior conditioning vector.

First, the phase-contrast timeline information, known a priori, is encoded into a unique 128-dimensional contrast phase prior, as shown in Fig. \ref{fig:Arc}(c). This prior is constructed such that its non-zero segment uniquely represents the contrast-phase timeline. Although it does not correspond to the actual elapsed time after contrast injection, the representation is designed to be non-overlapping across different phases, ensuring a distinct and interpretable encoding of temporal context. This prior serves as a key input to the Prior Conditioning Module (PCM).
The PCM integrates this contrast phase prior with encoder features from the Energy Conversion Module (ECM) to guide the decoding process, enabling effective adaptation to contrast dynamics across phases. Specifically, the contrast phase prior is first processed through a two-layer MLP-based Prior Embedding Block (PEB), producing a 128-dimensional phase-context embedding. In parallel, features from each encoder layer are aggregated using global average pooling (GAP) and transformed through a two-layer MLP to produce a 128-dimensional feature vector, providing spatial and structural context from the image that complements the phase prior. These two embeddings—the prior and the encoder feature—are concatenated and passed through an additional two-layer MLP to generate the prior conditioning vector (PCV). This fusion ensures that both temporal context (from the prior) and image-derived representations jointly influence the decoding process. This procedure is repeated across four PCMs (PCM1–PCM4), each associated with one of the four encoder layers of the ECM, as shown in Fig. \ref{fig:Arc}(b). Each PCM outputs a PCV of size 32, 64, 128, and 256, respectively, to align with the channel dimensions of the corresponding decoding layers.

In ECM, each encoder and decoder block contain two convolutional layers, each followed by batch normalization and ReLU activation. The prior conditioning vector is applied via channel-wise multiplication to the output of the second convolutional block in each decoder layer. This mechanism enables the network to adaptively modulate decoded features based on the input contrast phase, enhancing its ability to perform accurate, phase-specific energy transformation

\subsection{Data preparation and Training}
We utilized monochromatic image pairs at 70~keV and 50~keV from GE Healthcare's Gemstone Spectral Imaging (GSI) system for training the network. The dataset comprises 20 contrast-enhanced CT exams---five for each contrast phase: Angio, arterial, portal, and delayed. These Chest-Abdomen-Pelvis (CAP) scans were acquired using the GE Revolution Apex (GE Healthcare, Waukesha, WI) scanner with fast kVp-switching dual-energy acquisition (80~kVp and 140~kVp). Images were reconstructed using the Standard kernel with a slice thickness and spacing of 0.625~mm.

Out of the 20 volumes, 12 (three per contrast phase) were used for training and validation, and the remaining eight (two per phase) were reserved for testing. For each phase, 16{,}000 patches of size $128 \times 128$ were extracted from 2D axial images for training and 4{,}000 for validation. The model was trained for 600 epochs using the Adam optimizer with a batch size of 128. The initial learning rate was set to $1 \times 10^{-3}$ and decayed exponentially every third epoch with a gamma factor of 0.99. To balance perceptual quality and pixel-wise accuracy, the loss function combined Mean Absolute Error (MAE) with Structural Similarity Index Measure (SSIM), where SSIM was weighted by a factor of 0.05. 

The same architecture was employed for both unified and contrast-specific training, with the only difference being the inclusion of the Prior Conditioning Module in the unified model. The data used for each phase in unified and contrast-specific training were identical, and all other training parameters were kept consistent. All experiments were conducted on an NVIDIA RTX 6000 GPU with 50~GB of memory.

\section{Results}
In the first part of this section, we provide qualitative and quantitative comparison of the proposed method with stand-alone contrast specific training model outputs. In the later section, we have inferenced the model on actual 120kVp images from SECT system to demonstrate contrast improvement of 50keV like images generated by the proposed method across different contrast phases. 

\subsection{Comparison of Unified Model Performance with Stand-alone Models on DECT data}

\begin{figure}[htbp]
	\centering
	\includegraphics[width=0.57\textwidth]{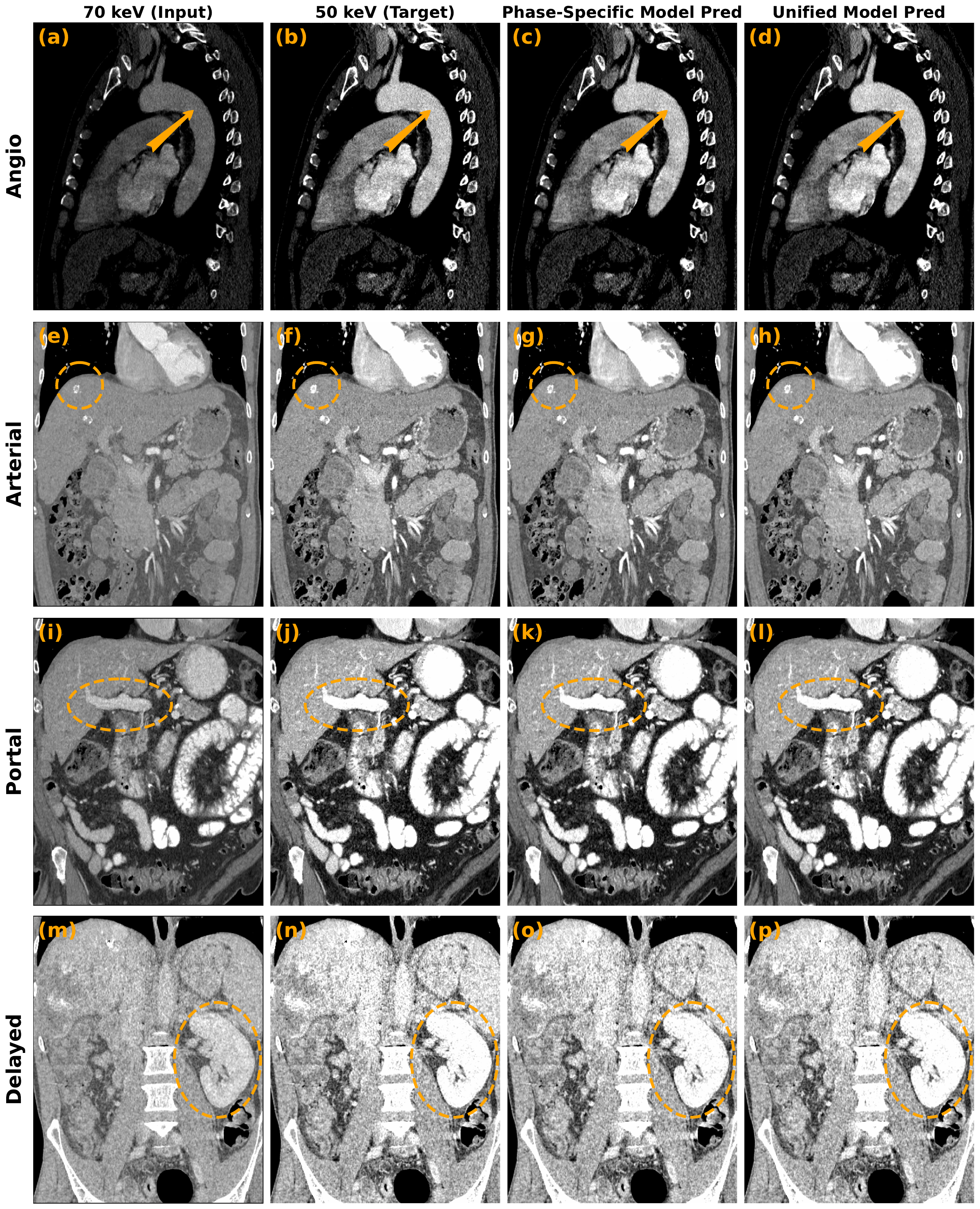}
	\caption{Illustration of contrast enhancement on DECT data (across columns) for Input 70keV, target 50keV, predicted 50keV from phase specific model and proposed unified framework respectively. (a) – (d) Angio phase; (e) – (h) Arterial phase; (i) – (l) Portal phase; (m) – (p) Delayed phase}
	\label{fig:DECT_Results}
\end{figure}

Fig. \ref{fig:DECT_Results} illustrates sagittal/coronal slices across four contrast phases—Angio, Arterial, Portal, and Delayed—comparing the input 70‑keV images, the target 50‑keV images, and predictions from phase‑specific models and unified model. For the Angio phase (Fig. \ref{fig:DECT_Results} a–d; WL = 300, WW = 600), the aorta is prominently enhanced on the 50‑keV target image relative to the 70‑keV input, and both model predictions produce same level of enhancement. Fig. \ref{fig:DECT_Results} e–h shows the Arterial phase (WL = 40 HU, WW = 800 HU), where contrast begins to perfuse into organs such as the liver and kidneys; similar enhancement of liver tumors is observed on the 50‑keV target and predicted images compared with the 70‑keV input.

For the Portal phase (Fig. \ref{fig:DECT_Results} i–l; WL = 100 HU, WW = 600 HU), peak enhancement of the liver and portal venous system is evident, which is crucial for detecting hypovascular lesions. Both prediction models effectively reproduce this enhancement. In the Delayed phase (Fig. \ref{fig:DECT_Results}, m–p; WL = 300, WW = 600), contrast washout occurs, yet residual enhancement persists in hepatic lesions and the renal parenchyma; again, both models closely match the target image.

The quantitative evaluation of the proposed models across different contrast phases is summarized in Table~\ref{tab:quant_results}. Metrics including Mean Absolute Error (MAE), Root Mean Square Error (RMSE), and Structural Similarity Index (SSIM) were computed for both the unified model and contrast-specific models. Interestingly, the unified model achieves slightly lower MAE and higher SSIM compared to contrast-specific models. This can be attributed to the joint training strategy, where all contrast phases are learned together with proper conditioning for each phase. Such multi-phase training enables cross-phase knowledge transfer, allowing the model to capture shared anatomical structures and intensity distributions across phases. This acts as an implicit regularizer, improving generalization and reducing overfitting to phase-specific noise.

\begin{table}[htbp]
	\centering
	\caption{Quantitative performance across contrast phases for Unified vs. Contrast-Specific models.}
	\label{tab:quant_results}
	\small
	\begin{tabular}{l
			r@{\,\,$\pm$\,\,}l
			r@{\,\,$\pm$\,\,}l
			r@{\,\,$\pm$\,\,}l
			r@{\,\,$\pm$\,\,}l
			r@{\,\,$\pm$\,\,}l
			r@{\,\,$\pm$\,\,}l}
		\toprule
		\multicolumn{1}{c}{\textbf{Phase}} &
		\multicolumn{4}{c}{\textbf{MAE (HU)}} &
		\multicolumn{4}{c}{\textbf{RMSE (HU)}} &
		\multicolumn{4}{c}{\textbf{SSIM}} \\
		\cmidrule(lr){2-5}
		\cmidrule(lr){6-9}
		\cmidrule(lr){10-13}
		& \multicolumn{2}{c}{Unified} & \multicolumn{2}{c}{Contrast-Specific}
		& \multicolumn{2}{c}{Unified} & \multicolumn{2}{c}{Contrast-Specific}
		& \multicolumn{2}{c}{Unified} & \multicolumn{2}{c}{Contrast-Specific} \\
		\midrule
		Angio      & 8.85 & 2.13 & 9.58 & 1.98 & 17.82 & 7.42 & 17.38 & 6.42 & 0.9995 & 0.0005 & 0.9994 & 0.0005 \\
		Arterial & 7.75 & 1.11 & 8.85 & 1.27 & 14.38 & 2.99 & 17.97 & 3.81 & 0.9992 & 0.0005 & 0.9989 & 0.0006 \\
		Delayed  & 6.81 & 0.88 & 7.23 & 0.94 & 11.97 & 1.74 & 12.75 & 1.78 & 0.9994 & 0.0001 & 0.9993 & 0.0001 \\
		Portal   & 7.43 & 1.21 & 7.60 & 1.83 & 13.71 & 3.36 & 14.54 & 6.49 & 0.9975 & 0.0006 & 0.9976 & 0.0005 \\
		\bottomrule
	\end{tabular}
\end{table}

\subsection{50keV like image generation from SECT Data}

\begin{figure}[htbp]
	\centering
	\includegraphics[width=0.48\textwidth]{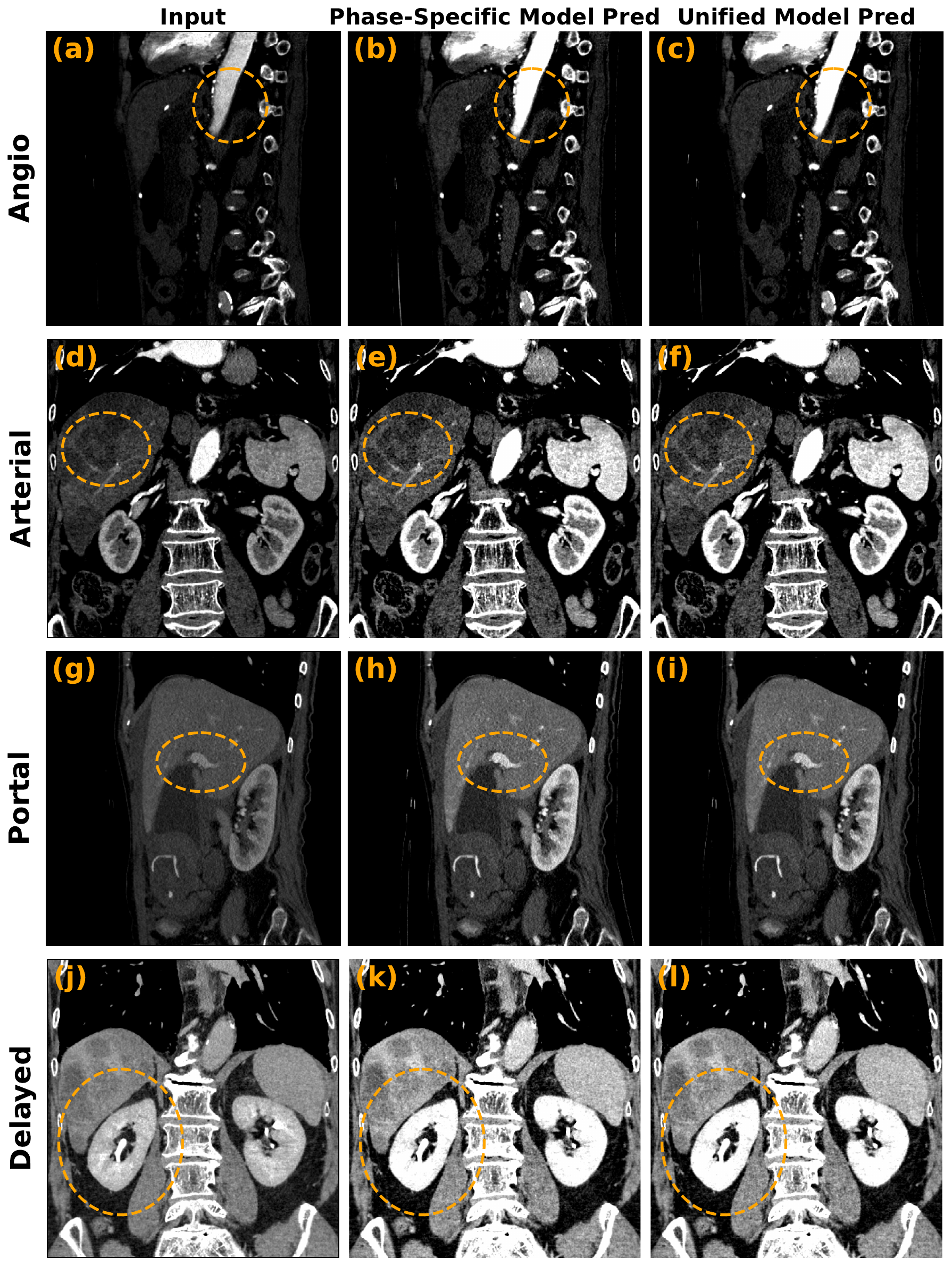}
	\caption{Illustration of contrast enhancement on SECT data (across columns) for Input 120kVp SECT, predicted 50keV from contrast specific model and proposed unified framework respectively. (a) – (c) Angio phase; (d) – (f) Arterial phase; (g) – (i) Portal phase; (j) – (l) Delayed phase}
	\label{fig:SECT_Results}
\end{figure}

In this section, we present 50~keV–like images generated using the proposed framework for the 120~kVp SECT contrast CT data acquired on a GE Revolution Ascend system, reconstructed using the Standard kernel with a slice thickness and spacing of 0.625~mm, and compare them with standalone prediction outputs. Fig.~\ref{fig:SECT_Results} shows the images of the 120~kVp data and the predicted images from the unified framework and phase-specific trained models across all four contrast phases. Similar to the contrast-phase images shown in Fig.~\ref{fig:DECT_Results}, the contrast enhancement by the proposed framework follows phase-specific contrast dynamics, highlighting anatomies according to the contrast phase timeline.

Fig.~\ref{fig:SECT_Results} a–c presents the Angio phase (WL = 300~HU, WW = 600~HU), where the aorta exhibits marked enhancement in the 50~keV–like images generated by both the proposed unified model and standalone methods, compared to the conventional 120~kVp input. In the Arterial phase (Fig.~\ref{fig:SECT_Results} d–f; WL = 40~HU, WW = 800~HU), liver tumours are more conspicuously visualized with improved contrast in the output images. The Portal phase (Fig.~\ref{fig:SECT_Results} g–i; WL = 100~HU, WW = 600~HU) demonstrates peak enhancement of the liver and portal venous system in the 50~keV reconstructions. In the Delayed phase (Fig.~\ref{fig:SECT_Results} j–l; WL = 300~HU, WW = 600~HU), residual enhancement is evident in both hepatic lesions and the renal parenchyma.

Comparing the images, the unified model consistently delivers enhanced, phase-appropriate contrast across all imaging phases, outperforming the 120~kVp baseline and achieving performance on par with the individually trained phase-specific models.

\subsection{Analysis of Phase-Conditioning Vectors}

The modulation vectors PCV1, PCV2, PCV3, and PCV4 illustrated in Fig.~\ref{fig:Arc}(a), have dimensions of 32, 64, 128, and 256, respectively. These vectors are designed to modulate the features of the decoder layer, enabling effective energy transformation across multiple contrast phases. Each vector contributes uniquely to feature modulation, ensuring that the network adapts to the varying characteristics of different phases.

\begin{figure}[htbp]
	\centering
	\includegraphics[width=0.50\textwidth]{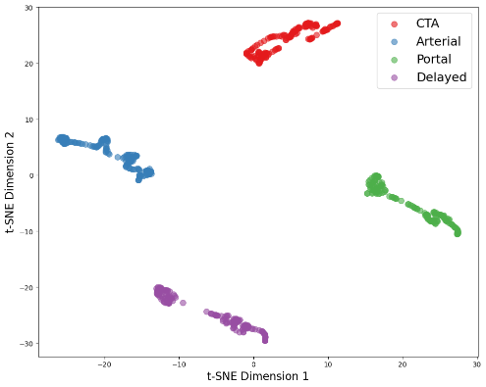}
	\caption{\textit{t-SNE} visualization of phase conditioning vectors. Distinct clusters correspond to CTA, Arterial, Portal, and Delayed phases.}
	\label{fig:tsne}
\end{figure}

To further analyze this functionality, all conditioning vectors (PCVs) are concatenated into a single vector of length 480. This combined representation is then projected onto a two-dimensional plane using \textit{t-SNE} visualization \cite{maaten2008visualizing}, as shown in Fig.~\ref{fig:tsne}. The t-SNE plot demonstrates clear separation among the modulation vectors corresponding to each contrast phase, indicating that the vectors effectively capture phase-specific characteristics.

\subsection{Computation comparison of Standalone and proposed framework }
Table 1 presents a comparative analysis of performance metrics between individual Stand-alone Models and the Proposed Unified Framework, evaluated on an Nvidia RTX 6000 50 GB GPU using 512×512 resolution images. The unified model demonstrates only a marginal increase in computational requirements, with parameters, model size and inference time, making it suitable for clinical deployment.

\begin{table}[ht]
	\centering
	\caption{Comparison of Stand alone vs Unified and Overhead}
	\label{tab:standalone_unified_overhead}
	\begin{tabular}{lccc}
		\toprule
		& \textbf{Stand alone} & \textbf{Unified} & \textbf{Overhead} \\
		\midrule
		Parameters        & 7.849 M  & 8.561 M  & 0.712 M (9.1\%)   \\
		Model Size        & 29.9 MB  & 32.7 MB  & 2.72 MB (9.1\%)   \\
		FLOPs             & 55.906 G & 55.922 G & 0.016 G (0.02\%)  \\
		Peak Memory Usage & 235.9 MB & 276.6 MB & 40.71 MB (17.3\%) \\
		Inference Time    & 3.35 ms  & 3.48 ms  & 0.14 ms (4.1\%)   \\
		\bottomrule
	\end{tabular}
\end{table}

\section{CONCLUSIONS}
We introduce a unified deep learning framework for generating virtual monochromatic images from SECT data, conditioned on contrast phase information. A novel Phase Conditioning Module (PCM) integrates this prior into the energy conversion network, guiding phase-specific energy transformation. Trained on DECT-derived data and evaluated across four contrast phases, the unified model matches the performance of phase-specific models in both visual quality and HU accuracy. It generalizes effectively to SECT inputs, producing 50 keV-like images with appropriate phase-specific contrast. This approach bridges the gap between SECT and DECT, offering spectral imaging benefits with simplified deployment and maintenance by eliminating the need for multiple models, reducing operational complexity in clinical workflows.

\bibliography{report} 
\bibliographystyle{spiebib} 

\end{document}